# NOMA for Next-generation Massive IoT: Performance Potential and Technology Directions

Yifei Yuan, Sen Wang, China Mobile
Yongpeng Wu, Shanghai Jiaotong University
H. Vincent Poor, Princeton University
Zhiguo Ding, University of Manchester
Xiaohu You, Southeast University（China）
Lajos Hanzo, University of Southampton

## ABSTRACT

**Broader applications of the Internet of Things (IoT) are expected in the forthcoming 6G system, although massive IoT is already a key scenario in 5G, predominantly relying on physical layer solutions inherited from 4G LTE and primarily using orthogonal multiple access (OMA). In 6G IoT, supporting a massive number of connections will be required for diverse services of the vertical sectors, prompting fundamental studies on how to improve the spectral efficiency of the system. One of the key enabling technologies is non-orthogonal multiple access (NOMA). This paper consists of two parts. In the first part, finite block length theory and the diversity order of multi-user systems will be used to show the significant potential of NOMA compared to traditional OMA. The supremacy of NOMA over OMA is particularly pronounced for asynchronous contention-based systems relying on imperfect link adaptation, which are commonly assumed for massive IoT systems. To approach these performance bounds, in the second part of the paper, several promising technology directions are proposed for 6G massive IoT, including linear spreading, joint spreading & modulation, multi-user channel coding in the context of various techniques for practical uncoordinated transmissions, cell-free operations, etc., from the perspective of NOMA.**

## I. INTRODUCTION

It is envisioned that 10 million connections per square kilometer will be required in 6G, which is ten times higher than the corresponding key performance indicator (KPI) of massive machine-type communications (mMTC) in 5G. This should spur wide participation by both academia and industry for seeking practical enabling solutions to support such demanding massive connectivity.

The development of the Internet of Things (IoT) in the 3$^{rd}$ Generation Partner Project (3GPP) dates back to 2011, when a simplified version of the 4G Long-Term-Evolution (LTE) standard was specified to support low-cost devices. Subsequently, narrowband IoT (NB-IoT) and enhanced machine-type communication (eMTC) were specified in 3GPP. Although originally designed for 4G, the NB-IoT and eMTC radio access technologies were transplanted into the 5G network architecture to fulfill the International Mobile Telecommunication (IMT)-2020 requirements.

At the beginning of the 5G standardization, non-orthogonal multiple access (NOMA) was considered as a promising solution for mMTC. This is because NOMA, when accompanied by grant-free transmissions, can improve the connection density and reduce the system overhead. However, the standardization of NB-IoT profoundly affected the NOMA as well as 5G mMTC developments in 3GPP: the low power wide-area (LPWA) scenario, an important aspect of mMTC, was removed from consideration to avoid the overlapping with NB-IoT.

Another issue is that due to the acceleration of the New Radio (NR) specification in 3GPP to complete the basic physical layer functions, the study on advanced features such as NOMA were postponed to Rel-16 [1], leaving limited room and time for fundamental changes to exploit the potential gain of NOMA. Consequently, no NOMA-based solution has been specified in 5G to support mMTC.

Over the last few years, NOMA assisted massive IoT has been a hot research topic [2]. From an information theoretic viewpoint, the fundamental potential of NOMA in supporting massive IoT was analyzed in [3], where the preliminary bounds were derived to show a significant gain for NOMA over orthogonal multiple access (OMA). Channel coding or NOMA is also emerging [4-5], which enriches the study in both fields of channel coding and signal processing. Multi-dimensional modulation [6] continues to capture some research interest. The performance of NOMA was also analyzed in [7]. Data-only based channel estimation was proposed in [8].

Research on 6G is still at its early stages and should embrace bold innovative designs for massive IoTs. In this paper we outline the ultimate performance goals of massive IoT and highlight several technology directions related to NOMA. It should also be noted that some IoT devices may be connected to repeaters or relays, where the associated channels can be quite different from the conventional models. Interested readers may refer to [9].

The paper is organized as follows. In Section II, a general massive IoT setting is discussed, where the performance bounds of NOMA assisted massive IoT are described Section III is devoted to the technology directions that are important for 6G massive IoT. Finally, the paper is concluded by Section IV.

## II. PERFORMANCE POTENTIAL

4G and 5G cellular networks are primarily scheduling-based, where link adaptation closely tracks the channel state information (CSI) and selects a suitable transmission format. The system performance is usually evaluated in terms of data rates, often estimated by using the Shannon capacity for each link assuming infinitely long blocks. This assumption may be appropriate for enhanced mobile broadband (eMBB) services whose packet size is often large.

By contrast, for massive IoT, the typical traffic consists of infrequently-sent small packets, e.g., 40~320 bits for each packet, where Shannon's asymptotic capacity formula becomes invalid.

In massive IoT scenarios, link adaptation becomes a burden, simply due to the excessive control/signaling overhead for handshaking, CSI estimation, feedback, scheduling grants, etc., compared to the small payload. The link adaptation has to operate in a radio resource control (RRC) connected mode, where the multi-step random access procedure, including random access requests and responses, User ID reports and connection set-up, scheduling requests and UL scheduling grant transmissions, should be carried out between a base station (BS) and a terminal device (blue and red arrows represent the downlink and uplink, respectively) before the terminal can transmit data, as illustrated in Figure 1. As a result, hundreds of bits have to be exchanged between the terminal and the BS, although the terminal might just have to send a much smaller number of data bits. This motivates grant-free/autonomous transmission as illustrated in Figure 1 where the terminal can immediately send the data packet.

Grant-free access refers to the pre-allocation of signatures for users to avoid the potential collisions. Such mechanism may be suitable for stationary terminal devices that seldom perform cell switching. For grant-free access, NOMA can approach the system capacity, but OMA generally cannot.

Autonomous transmission implies contention-based access: users may select the same signatures. This is a more general use case for massive IoT in which pre-allocations are often impractical and the closed-loop power control is not feasible either.

In grant-free or autonomous transmissions, the system performance should be evaluated in terms of number of users supported, under a certain target per-user rate and with certain outage probability. An outage event is declared when the data rate supported by the instantaneous fading channel is below the target data rate.

For massive IoT, the operation may resemble random access when a radio link is not fully established. The access mechanism can be contention-free (e.g. pre-configured) or contention-based (e.g. uncoordinated), where two or more users may choose the same signature for preamble transmission in the latter case. The signatures may serve as temporary identifiers (IDs) and also be used for channel estimation. Signatures can take the forms of reference signals, spreading codes, interleaver patterns, etc. The number of signatures should be set appropriately to: 1) avoid the excessive number of blind detections; 2) ensure a low collision probability. Note that the chance of two users transmitting exactly the same set of bits is extremely low even for short packets. If the user ID is carried as part of the payload, the difference between contention-based and contention-free access becomes blurred from a physical layer perspective. Hence, the performance analysis of Section II is generally applicable to both.

In the remainder of this section, we will first discuss achievable performance bounds for NOMA with short packets mainly for Rayleigh faded channels, by using the outage probability as the metric. In the second sub-section, the performance potential of NOMA is quantified by its multi-user diversity gain. The third sub-section is about the synergy of the two approaches.

### A. Performance bounds of a multiple access channel having short block length

Since Shannon established the capacity formula for a point-to-point link, various rate bounds have been proposed. The rate bounds derived in [10] are significantly tighter compared to many previous estimates derived for blocklengths as short as 100 bits. Normal distribution is used for drastically simplifying the bound equations. It turns out that the maximum achievable rate can be represented as a certain fraction of the asymptotic Shannon capacity, which is a function of the channel dispersion, the block length and the error probability. The maximum achievable rate decreases as the channel dispersion increases, the blocklength gets shorter, or the error probability is set lower.

While the maximum achievable rate derived for a single-user link is useful for OMA, it cannot readily be used for NOMA that can approach the performance bounds of multiple access channels. The information-theoretic study of multiple access channels is a relatively new research area where many problems are still open. A simple form of the multiple access channel is the Gaussian multiple access channel (GMAC), where the signal to noise ratios (SNRs) are the same and constant for all users, e.g., the channel only suffers from AWGN. Several

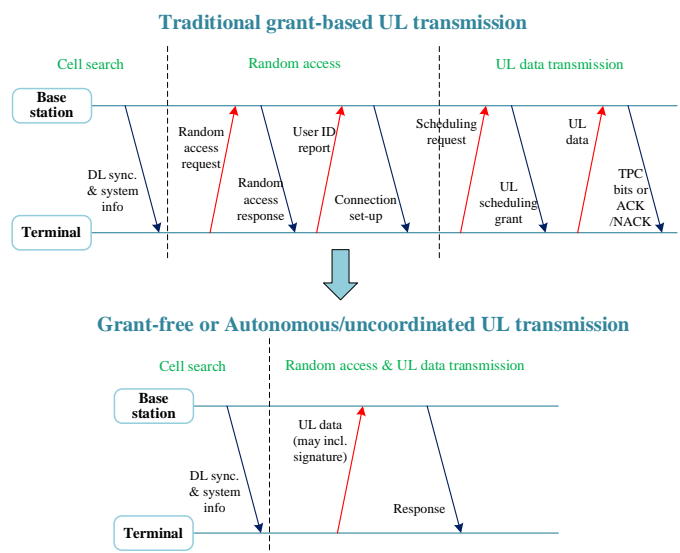

**Figure 1** *Grant-based vs. grant-free or autonomous /uncoordinated uplink (UL) transmission*

performance bounds have been derived for the GMAC having finite blocklengths.

For massive IoT, the GMAC of a static channel is not general enough since only open-loop power control would be feasible to compensate the pathloss and shadow fading. In the absence of closed-loop power control (due to the signaling overhead constraint of massive IoT), the received SNR of each user would fluctuate with the fast fading. In this sense, the multiple access channel experiencing independent Rayleigh fading per user is more appropriate for massive IoT scenarios. In the literature, very few bounds have been developed for multi-user fading channels of short blocklengths. In the insightful analysis of [3], the performance bounds of Rayleigh faded MAC are derived and compared to that of orthogonal e.g., time-division multiple access (TDMA). Since NOMA is deemed to be able to approach MAC bounds, these preliminary new bounds reveal the promising potential of NOMA for massive IoT. By substituting the relevant parameters into the analysis of [3], Figure 2 is plotted for the Rayleigh faded MAC, where the system load $\mu$ is the ratio between the number of users $K$ and the total number of resources $n$. Here, the massive IoT is modeled by assuming that both $K$ and $n$ are infinitely large (while $\mu$ is still finite). Those performance bounds associated with perfect CSI and no CSI knowledge at the receiver are denoted as CSIR and no-CSI, respectively in Figure 2, Here, the blocklength $k$ is assumed 100 bits, which is typical for massive IoT traffic. The product of the system load and the information block size represents the total spectral efficiency of the system. The target per-user outage probability (also called individual outage probability) is 0.01.

The basic findings from Figure 2 are that when the system load $\mu$ or the system's spectral efficiency $k\mu$ is low, the required energy per bit vs. power spectral density of noise (Eb/No) for NOMA and OMA (e.g. TDMA) are very similar. As the system load increases, the total Eb/No required for NOMA remains almost the same, for both the upper bounds and the lower bounds. This means that in this "cliff" or "hysteresis" region, adding more users, consequently having more interference, would not require higher total SNR (Eb/No). By contrast, for OMA, the total Eb/No increases significantly when the system load increases. In particular, when the system load reaches the top of the "cliff" region (e.g., $\mu = 0.16$, system spectral efficiency of 16 bps/Hz), the performance gap between NOMA and OMA widens to over 20 dB. As the system load further increases, the Eb/No required for NOMA increases at a pace similar to that of OMA.

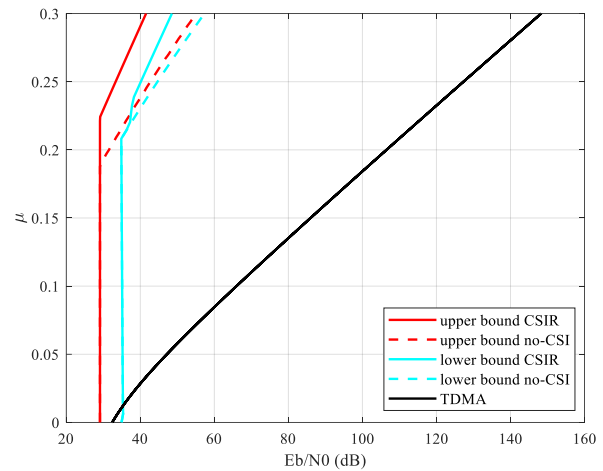

**Figure 2** *The system load $\mu$ vs. total Eb/No for Rayleigh fading channels in massive random access. Information block length k = 100 bits*

### B. Upper bound of multi-user diversity gain

For massive IoT, the performance gain of NOMA over OMA can also be represented by its multi-user diversity [7] which is inherent to NOMA, but somehow has not attracted enough attention. Here, the multi-user diversity gain is analyzed assuming that the sum rate of multiple users can be represented using classic Shannon capacity analysis.

Again, the individual outage probability (OP) is used, which is derived from the joint OP, Explicitly, a joint outage event is declared when the sum rate of a subset of users is below a predefined target sum rate. The individual outage probability of a user (mentioned in Section II-A) is the sum of the joint OPs over all the combinations of the subsets of users that contain this user. The derivation assumes joint detection and decoding at the receiver. These conditions represent a performance upper bound of NOMA for uplink grant-free transmissions. Similar to [7], the individual OPs as functions of the SNRs for Rayleigh faded MAC are calculated via Monte-Carlo simulations. Some results are presented in Figure 3 with various numbers of users (denoted as $K$) sharing the same time-frequency resources whose $K = 1$ represents the OMA case. The multiuser diversity here is defined as the total SNR gain of NOMA over OMA at a specific individual OP of certain sum rate. A major finding is that the multi-user diversity gain of NOMA increases as more users participate in NOMA transmissions. This gain saturates once the number of users reaches a certain value, for example, 4 users and 8 users as observed for the sum rate of 3 bps/Hz and 9 bps/Hz, respectively. The multi-user diversity gain also increases as the sum rate increases to 9 bps/Hz. A more substantial gap is observed between "K=1, $R_{sum}$= 9" and "K=16, $R_{sum}$=9" than between "K=1, $R_{sum}$=3" and "K=1, $R_{sum}$=9".

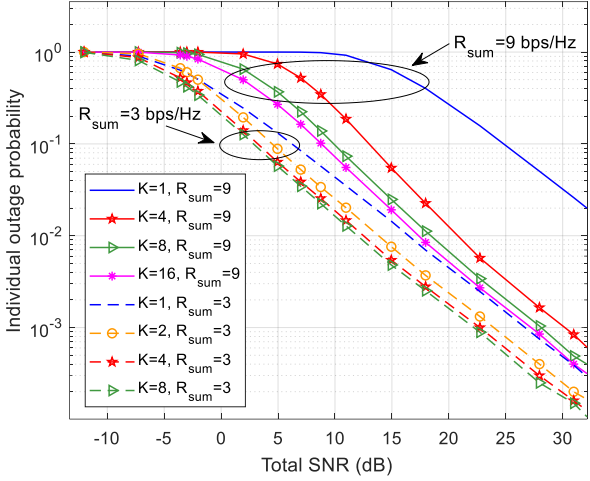

**Figure 3** *Individual user outage probability vs. total SNR for the independently Rayleigh faded multiple access channel*

### C. Synergy of information-theoretic study and multi-user diversity analysis

Even though the two approaches discussed in II-A and II-B are very different, it turns out that the findings from Figure 3 are quite consistent with those from Figure 2. Both show that when the number of superimposed users increases, the performance gain of NOMA over OMA increases. The range of system spectral efficiency of interest is also similar, e.g., 3~16 bps/Hz between Figure 2 and Figure 3.

The information-theoretic approach considers finite block lengths and it is particularly insightful for small packet transmission in massive IoT. Nevertheless, those bound calculations require asymptotic treatment, which for example both the number of users and the number of resources go to infinity. On the other hand, while the multi-user diversity approach conveniently estimates the performance bound for finite numbers of users (e.g., saturation points) and for various types of receivers including the suboptimal successive interference cancellation (SIC), the effect of finite blocklength cannot be reflected.

It is expected that by jointly considering the impact of short blocklengths and multi-user diversity, more industrially relevant performance bounds can be derived, which may provide more guidance and insight into the practical design of NOMA for massive IoT. In this regard, [11] is an initial attempt to explore this synergy.
.

## III. TECHNOLOGY DIRECTIONS

The above discussions show that for massive IoT relying on grant-free/autonomous transmissions, NOMA can potentially offer much higher spectral efficiency than OMA. Although the NOMA technology has been studied in 5G, particularly in 3GPP, the understanding of NOMA for uplink massive IoT is still far from being deep and comprehensive. The first open problem is the ultimate system performance where only recently some initial analysis has emerged (as shown in Section II). It is believed that these theoretical studies may provide further guidance on practical solutions. Technology wise, the primary goal is to design the physical layer signal format to approach the performance bounds in Section II at the system level, while being able to operate in uncoordinated transmission environment with reasonable complexity of receivers.

The major challenge in signal format design is how to "spread" the information bits to facilitate the decoupling of bits between users at reasonable receiver complexities. The general sense of "spreading" essentially maps a relatively short sequence of information bits of a user to a longer sequence of modulated symbols, which can be achieved by symbol-level linear spreading, joint modulation and spreading, or the emerging multi-user channel coding (suggested by the theoretical analysis in Section II) as illustrated in Figure 4.

To fulfill this performance potential of NOMA for massive IoT, particularly at the system level, the other cell interference should be handled properly, e.g., either to be suppressed or considered part of the signals to be detected. In this sense, cell-free operation would be crucial

Uncoordinated transmissions pose quite a lot of design challenges which tend to be more pronounced when combined with NOMA. This encourages new reference signal designs, new waveforms, efficient receiver algorithms, frame structures, etc.

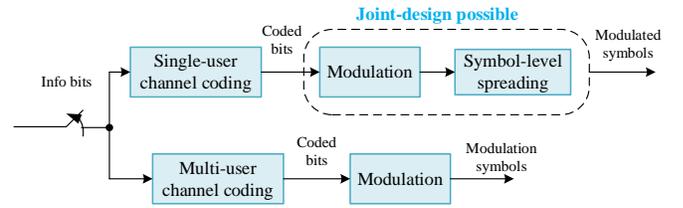

**Figure 4** *Single-user channel code vs. multi-user channel code based uplink NOMA transmitter for massive IoT*

### A. Signal spreading based schemes

Signal spreading based schemes, as shown in the upper branch of Figure 4, have been studied in 3GPP [1], [12] and can be divided into two groups: 1) symbol-level linear spreading; 2) joint modulation and spreading.

Symbol-level linear spreading directly applies a spreading sequence to each modulated symbol. The word "linear" emphasizes that traditional modulation, e.g. quadrature phase-shift keying (QPSK), would be used. The actual sequences for linear spreading can be designed under the different cross-correlation criteria, which differentiates the schemes proposed in 3GPP. For massive IoT, , the design should also ensure large pools of sequences to avoid potential collisions in autonomous uplink transmissions.

The typical receivers designed for linear spreading rely on minimum mean-square error (MMSE) de-spreaders using interference cancellation (IC). Hard-decoded bits are used for regenerating the interference signal for cancellation. MMSE-hard IC has a relatively low complexity and its architecture is compatible with the receiver implementations in typical 4G and 5G base stations. However, the hard-IC receiver is suboptimal and its achievable rate is lower than that of the optimal joint detector. From the per-user outage perspective, its performance gap to joint detection increases as the number of users or/and per user spectral efficiency increases.

In joint modulation and spreading, the encoded bits are directly mapped to a series of (spread) modulated symbols, as seen in the dashed box of Figure 4. It is also called multi-dimensional spreading, which can exploit the extra diversity gain of the joint modulation constellations of different users. To maximize the "shaping" gain, the constellations would deviate from traditional QPSK or quadrature amplitude modulation (QAM) constellations. Sparsity can be introduced to reduce the complexity of the receiver implementation, as in sparse code multiple access (SCMA). Note that traditional single-user channel codes are still used here.

This joint design may have an improved performance. However, a high-complexity maximum-likelihood receiver is required, although the complexity can be reduced to some extent by using sparse codes. The joint design also has a scalability issue: its codebooks are specifically customized for certain spreading lengths and have to be optimized separately for different numbers of users, which can be problematic for massive IoT.

### B. Multi-user channel coding

Compared to the signal processing type of approaches in Section III-A, more fundamental solutions would include new channel codes, as illustrated in the lower branch of Figure 4. The reason is two-fold. Firstly, the data packets in massive IoT are typically small, whereas traditional channel codes designed for broadband traffic are optimized for large blocklengths; Secondly, channel coding used for wireless communications typically targets the single-user case, because the predominately orthogonal multiple access of legacy systems reduces the need for multi-user codes. As suggested in Section II, the multi-user performance bounds show a significant potential for performance improvements using more advanced channel codes, presumably of the multi-user type. This potential approach is indeed plausible, because for the single user case (e.g. OMA) the channel dispersion [10] becomes significant for short blocklength even in the AWGN channel. For fading channels, the situation is expected to get worse and the rate reduction compared to the infinitely long blocks would become even more grave. By using multi-user channel coding, together with more advanced receivers, substantial multi-user diversity gain can be gleaned in the scenario of fading channels and having short blocklengths.

Several practical multi-user code designs have already been conceived, for instance, a combination of repetition and low density parity check (LDPC) codes [4], spatially-coupled LDPC codes [5], etc. The basic idea for these multi-user codes is to introduce extra degrees of design freedom, e.g., repetition patterns, or coupling the Tanner graphs using different parameters in order to improve the robustness of the channel codes under more complex channels expected for uncoordinated multi-user systems supporting massive numbers of IoT devices. The candidate channel codes should not be limited to LDPC codes. More structured codes like Polar codes and other block codes can potentially be considered where for example, designs based on channel polarization can be generalized to exploit the significant near-far effect in massive IoT. Traditional channel code design methods, such as density evolution, EXIT charts, code distance analysis, etc., can be employed to hand-craft the channel codes for NOMA. Recently advances in artificial intelligence (AI) and machine learning have been found to be quite useful for designing good codes. In [13], it was proposed to use a genetic algorithm for constructing the parity check matrix of LDPC codes having short blocklengths, which exactly fits the small packet traffic pattern of massive IoT. In such a design approach, edges in the LDPC's Tanner graph are added or pruned via genetic trials, which departs from the classical design commencing from the optimization of the degree distribution of the parity-check matrices. With the pervasive applications of AI to various fields, it may become an indispensable tool for channel codes design.

MAC receiver implementation is also crucial for fulfilling the potential of multi-user channel code designs. The receiver algorithm is likely to be iterative, e.g., supporting extrinsic information exchange between the detector and the decoder. The decoder generates soft bits for joint iterative detection and decoding, but it can also perform joint decoding across different users for exploiting the multi-user code structures, while aiming for a low complexity.

### C. Cell-free architectures for massive IoT

The cell-free concept is particularly important in NOMA assisted massive IoT. It is well known that the inter-cell interference is a key limiting factor of system capacity. In NOMA systems, this factor becomes more limiting. As seen from Figure 2, the total SNR should be at least 20 dB for supporting massive IoT using NOMA. For typical multi-cell systems, the OPs are the highest for cell edge users, whose SNRs can rarely reach 20 dB. This poses a major challenge in exploiting the potential of NOMA at the system level. For NOMA solutions such as linear spreading, the inter-cell interference can be suppressed to some extent by employing a spreading code domain MMSE detector. This is possible because the covariance matrix of inter-cell interference exhibits a certain structure in the spreading code domain, owing to the limited number of dominant interfering signals from adjacent cells. The structure of interference can be exploited by the MMSE detector [1]. Despite the above benefit of the spreading code domain MMSE detector, the seamless integration of multi-user channel coding with interference suppression is an open challenge.

In the cell-free concept illustrated in Figure 5, a group of BSs would support a group of devices, which has the benefit of load-balancing and interference mitigation. This is in contrast to the traditional cell-based access, where each base station only responds to the devices within its cell boundary, which is also shown in Figure 5.

When the cell-free concept is applied to massive IoT, new challenges arise. For example, new random access procedures are required which can significantly differ from the traditional single-cell based random access. For instance, multiple BSs rather than one BS would perform the preamble detection and related control-plane processing for a user. The scenario is also quite different from the data-plane uplink coordinated multi-point (CoMP) philosophy that is typically scheduled by the BSs in RRC connected mode. This would have significant impact on the control plane designs of the protocol layers in next-generation systems.

### D. Handling practical issues in uncoordinated transmission

Again, in practical systems there are numerous challenges in supporting uncoordinated transmissions. One of them is how to carry out channel estimation for coherent detection. Typically, reference signals are used for this purpose and can be distinguished by orthogonal physical resources or codes. Given the limited number of orthogonal reference signals, collisions are likely in uncoordinated systems supporting hundreds of users simultaneously. To increase the pool size of reference signals, non-orthogonal reference signals may be considered, although this would degrade the performance. Advanced channel estimation contrived with iterative interference cancellation may be used for mitigating this issue. Data-only transmission mechanisms without reference signals [8] offer another alternative. The payload detection or demodulation can still be coherent, assuming that the channel coefficients can be directly estimated from the received payload, based on the geometric characteristics of the constellations of received signals. The resultant channel estimation can be accurate enough for low-order modulations such as binary phase-shift keying (BPSK) or QPSK. The receiver complexity would be increased due to the extra blind decoding/detection, compared to the case when the reference signal is known.

Massive IoT traffic often exhibits sparsity in the sense that, while the total number of devices is huge, the number of active devices is much smaller. This motivates the use of a compressed sensing-based approach for joint device activity detection and channel estimation. In [14] it is shown that the approximate message passing (AMP) algorithm can be extended to temporally or spatially correlated channels.

For uncoordinated transmissions without closed-loop timing adjustment to compensate for the different propagation delays, it is difficult for BS receivers to maintain precise time/frequency synchronization with the users. Suitable waveforms and frame structures should be designed, for example, to use narrower sub-carrier spacing for low mobility scenarios to increase the cyclic prefix duration with greater tolerance to the timing errors or differences. The frame structure can be based on slotted-Aloha to fit the asynchronous and collision-prone nature of massive IoT transmissions.

MIMO technology remains indispensable for next-generation communication systems, and thus NOMA and MIMO should work jointly. Fundamental studies indicate that if the number of receive antennas is large and the received power per degree-of-freedom is sufficient, SIC receiver can ensure very low error probability. However, in massive IoT scenarios, to support wide coverage, the carrier frequency may not be high. Hence, the number of antennas at the BS would be limited due to the relatively large size of antennas.

## IV. Conclusions

In this paper, the potential of non-orthogonal multiple access designed for 6G massive IoT has been analyzed in terms of its performance bounds for short blocklength coding and multi-user diversity for Rayleigh faded multiple-access channels. Promising gains have been demonstrated for NOMA over OMA. Several enabling technology directions have been discussed. In addition to the technologies already considered for 5G, such as linear spreading plus joint modulation and spreading, new directions have been suggested, such as multi-user channel coding that may potentially approach the performance bounds of uncoordinated MAC systems. The cell-free concept of massive IoT also has a high performance potential when combined with NOMA. Some of the practical aspects of the uncoordinated transmissions and MIMO deployment have also been discussed, such as channel estimation, reference signal design, signature collision, time/frequency asynchronization, receiver algorithms, etc. which are important for designing the robust NOMA-aided massive access.

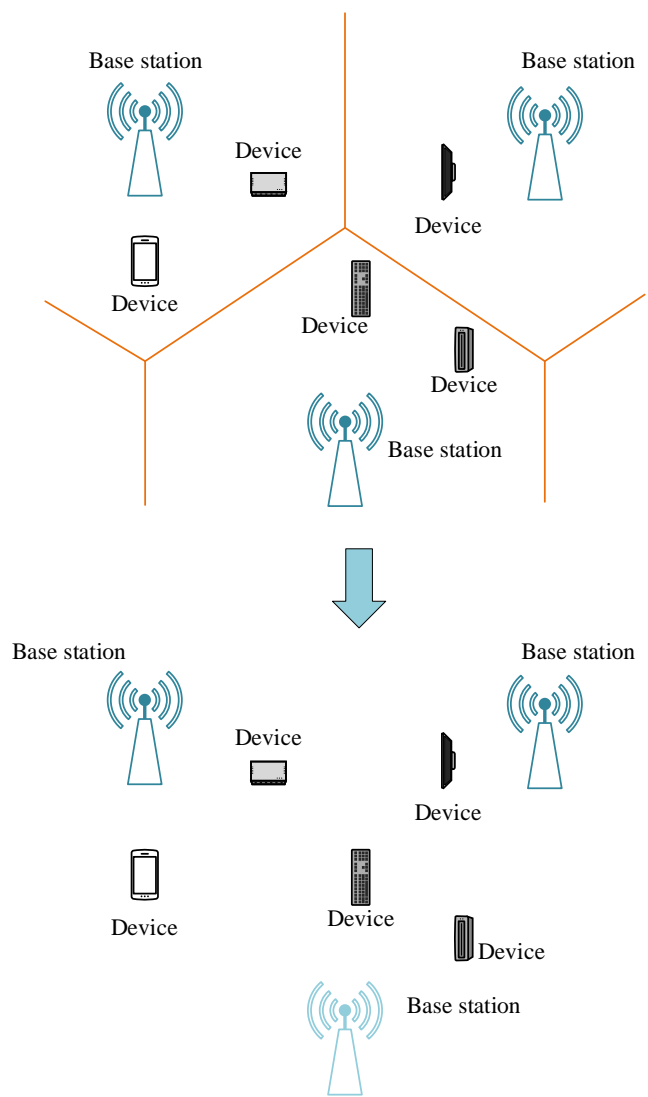

**Figure 5** *Cell-based vs. cell-free access for massive IoT*

# Biographies

Yifei Yuan (yuanyifei@chinamobile.com) was with Alcatel-Lucent from 2000 to 2008. From 2008 to 2020, he was with ZTE Corporation as a technical director and chief engineer, responsible for standards & research of LTE-Advanced and 5G technologies. He joined the China Mobile Research Institute in 2020 as a Chief Expert, responsible for 6G. He has extensive publications, including seven books on LTE-Advanced and 5G. He has over 60 granted patents.

Sen Wang is with China Mobile Research Institute. His research interests include optimization with application to wireless system, 5G/6G air interface technologies, especially on MIMO, waveform and multiple access. And he has published over 20 IEEE journal and conference papers in these areas. He holds 40 awarded and pending patents applications.

Yongpeng Wu is currently an associate professor at Shanghai Jiao Tong University, China. Previously, he was senior research fellow of Technical University of Munich and the Humboldt Research Fellow with Institute for Digital Communications, University Erlangen, Germany. His research interests include massive MIMO/MIMO systems, massive machine type communication, physical layer security, and signal processing for wireless communications.
.
H. Vincent Poor is with Princeton University. An IEEE Fellow, Dr. Poor is a member of the U.S. National Academy of Engineering and the U.S. National Academy of Sciences, and is a foreign member of the Chinese Academy of Sciences, the Royal Society and other national and international academies.  In 2017, he received the IEEE Alexander Graham Bell Medal.

Zhiguo Ding is currently a Professor at the University of Manchester. During Sept. 2012 to Sept. 2020, he was also an academic visitor at Princeton University. Dr Ding's research interests are 5G networks, signal processing and statistical signal processing. He has been serving as an Editor for IEEE TCOM, IEEE TVT, and served as an editor for IEEE WCL and IEEE CL. He is an IEEE Fellow and Web of Science Highly Cited Researcher.

Xiao-Hu You has been working with Southeast University, where he is currently the director of National Mobile Communications Research Laboratory. He has contributed over 300 IEEE journal papers. From 1999 to 2002, he was the Principal Expert of the C3G Project. From 2001-2006, he was the Principal Expert of the China National 863 4G FuTURE Project. From 2013 to 2019, he was the Principal Investigator of China National 863 5G Project

Lajos Hanzo (FIEEE'04, is a Fellow of the Royal Academy of Engineering. He was also awarded the Doctor of Sciences degree by the University of Southampton and Honorary Doctorates by the TU of Budapest and by the University of Edinburgh. He is a Foreign Member of the Hungarian Academy of Sciences and a former Editor-in-Chief of the IEEE Press. He has published 1900+ contributions at IEEE Xplore and 19 Wiley-IEEE Press books. He holds the Chair of Telecommunications at the University of Southampton.